\RequirePackage[colorlinks, linkcolor=blue, urlcolor=blue, citecolor=blue]{hyperref}
\RequirePackage[table,xcdraw]{xcolor}
\documentclass[draft]{agujournal2019}
\usepackage{url} 
\usepackage{lineno}
\usepackage[inline]{trackchanges} 
\usepackage{soul}

\usepackage{amssymb}
\usepackage{amsthm}
\usepackage{amsmath}
\usepackage{setspace}
\usepackage{listings}
\usepackage{color}
\usepackage{multirow}
\usepackage{url}

\newcommand{\be}{\begin{equation}}
\newcommand{\ee}{\end{equation}}
\newcommand{\bes}{\begin{equation*}}
\newcommand{\ees}{\end{equation*}}
\newcommand{\ba}{\begin{eqnarray}}
\newcommand{\ea}{\end{eqnarray}}
\newcommand{\bas}{\begin{eqnarray*}}
\newcommand{\eas}{\end{eqnarray*}}

\newcommand{\bfu}{{\bf u}}

\draftfalse

\journalname{Journal of Advances in Modeling Earth Systems (JAMES)}

\begin{document}

%
%

\title{Addressing out-of-sample issues in multi-layer convolutional neural-network parameterization of mesoscale eddies applied near coastlines}

%
%

\authors{Cheng Zhang\affil{1}, Pavel Perezhogin\affil{2}, Alistair Adcroft\affil{1}, Laure Zanna\affil{2}}

\affiliation{1}{Program in Atmospheric and Oceanic Sciences, Princeton University, Princeton, NJ 08542, USA}
\affiliation{2}{Courant Institute of Mathematical Sciences, New York University, New York, NY 10012, USA}


\correspondingauthor{Cheng Zhang}{cheng.zhang@princeton.edu}




\begin{keypoints}
\item This study validates specialized boundary condition treatments in CNN models to reduce boundary artifacts in ocean parameterizations.
\item This approach can be applied directly to already trained CNN models to ensure accurate and stable implementation of mesoscale eddies parameterizations.
\item Replicate padding outperforms zero padding by minimizing boundary artifacts and preventing extreme values that compromise simulations.
\end{keypoints}

%
%

%
%


\begin{abstract}
This study addresses the boundary artifacts in machine-learned (ML) parameterizations for ocean subgrid mesoscale momentum forcing, as identified in the online ML implementation from a previous study \cite{zhang2023online}.
We focus on the boundary condition (BC) treatment within the existing convolutional neural network (CNN) models and aim to mitigate the "out-of-sample" errors observed near complex coastal regions without developing new, complex network architectures.
Our approach leverages two established strategies for placing BCs in CNN models, namely zero and replicate padding.
Offline evaluations revealed that these padding strategies significantly reduce root mean squared error (RMSE) in coastal regions by limiting the dependence on random initialization of weights and restricting the range of out-of-sample predictions.
Further online evaluations suggest that replicate padding consistently reduces boundary artifacts across various retrained CNN models.
In contrast, zero padding sometimes intensifies artifacts in certain retrained models despite both strategies performing similarly in offline evaluations.
This study underscores the need for BC treatments in CNN models trained on open water data when predicting near-coastal subgrid forces in ML parameterizations.
The application of replicate padding, in particular, offers a robust strategy to minimize the propagation of extreme values that can contaminate computational models or cause simulations to fail.
Our findings provide insights for enhancing the accuracy and stability of ML parameterizations in the online implementation of ocean circulation models with coastlines.

\end{abstract}

\section*{Plain Language Summary}
This study focuses on improving machine learning (ML) models used to predict ocean forces near coastlines, where errors arise because these models lack information in the area. We investigated how boundary conditions are handled in existing convolutional neural network models to reduce these errors without creating complex new architectures. By using two methods, i.e., zero padding and replicate padding, we found that replicate padding significantly decreases prediction errors in coastal areas. While zero padding sometimes worsens issues in certain models, our results show that replicate padding is more reliable for effectively minimizing extreme value errors. This work highlights the importance of proper boundary condition treatment in ML models for coastal applications, ultimately aiming to enhance the accuracy and reliability of ocean circulation predictions.

%
%

\section{Introduction}\label{sec1}

Even with advances in computing over recent decades, climate models have finite resolution and must parameterize unresolved, subgrid-scale processes.
Historically, these parameterizations employ a mix of theory and empirical approaches \cite<e.g., for ocean circulation,>{gent_parameterizing_1995,griffies_isoneutral_1998,juricke2017stochastic}, but are imperfect so that the representation of subgrid processes continues to be a major source of bias and errors in climate projections \cite{stevens2013GCM,hewitt2020resolving}.

Recently, the use of machine learning methods has emerged as a promising tool for developing subgrid parameterizations in numerical models of both the atmosphere \cite{rasp2018deep,beucler2021climate,yuval2021GRL,wang2022nonlocal,shamekh2023precipitation} and ocean \cite{Bolton2019,zanna2020data,Guillaumin1&Zanna-JAMES21,sane2023verticalmixing,ross2023benchmarks,bodner2023submeso, perezhogin2023implementation}.
Among machine learning architectures employed for parameterization of subgrid fluxes in climate models, convolutional neural networks (CNNs) have become increasingly popular due to their ability to connect local fluxes or tendencies to spatially non-local features \cite<e.g.,>{Bolton2019,zanna2020data,Guillaumin1&Zanna-JAMES21,bodner2023submeso,gregory2024seaice}.
However, applying CNN-based ML parameterizations presents unique challenges in ocean circulation models which must account for complex boundary conditions and topographical features at the Earth’s surface.
Unlike atmospheric models, ocean models must manage dynamic interactions of coastal water with shorelines, where CNNs often struggle to make a prediction, because they operate by sliding fixed-size kernels across images (fields) to extract features.
This was highlighted in a recent study by \citeA{zhang2023online}, where significant boundary artifacts were observed when a CNN, trained to parameterize mesoscale eddies in the open ocean\cite{Guillaumin1&Zanna-JAMES21}, was employed near coasts in an ocean circulation model.
These artifacts are likely the "out of sample" problem in ML parameterizations, wherein a network trained on limited data (open ocean) might extrapolate poorly beyond its knowledge base (near the coast).
In their study, the out-of-sample predictions of the CNN applied near the boundaries lead to errors that can ultimately propagate across the entire computational domain as the circulation model evolves.

Training a CNN model with global data including the near-coast regions should solve the out-of-sample problem near boundaries.
However, this approach has significant challenges.
Consider training a CNN to represent an open ocean (deep water) physical process that is unresolved in a coarse ocean model.
Many of the eddy-resolving process-studies used to derive parameterizations of mesoscale turbulence do not consider eddy interaction with coasts, and so inherently may not rectify the parameterized physics and fluxes appropriately.
Secondly, the shallow depth near the coast might modify the spatial scales of the process requiring even finer resolution in the high-resolution simulation used to obtain training data which could be too costly \cite{Hallberg-2013}.
Thirdly, the training data near the coastline is limited because the majority of the grid points correspond to the open ocean.
Finally, the modification of the process near coastlines (e.g. different dynamics, scales, etc.) likely needs more sophisticated CNN models to capture the additional complexity (i.e either or both deeper and wider networks).
The machine learning models that we have seen developed so far to parameterize ocean processes are based either on data from idealized simulations \cite{Bolton2019,zanna2020data,ross2023benchmarks} or on regional data from open ocean areas in global General Circulation Model (GCM) simulations that exclude land points \cite{Guillaumin1&Zanna-JAMES21,bodner2023submeso}.

All this to say, training a new ML model including coastlines remains challenging, and alternative strategies to mitigate the out-of-sample issues near shorelines should be explored.
The border effect for CNNs has been extensively studied in image processing, where common remedies include filling values at the image edges, that is padding \cite{innamorati2018simplepadding,nguyen2019randompadding,huang2021trainpadding,yang2023randompadding}, rescaling the result of the convolution operation near the borders, that is partial convolution \cite{liu2018partialconvolution},  or changing the filter kernel near the borders \cite{leng2023paddingfree}.
A most straightforward approach to mitigate out-of-sample errors involves providing boundary values for land points, i.e., filling in appropriate values where no ocean field data exists.
In previous research, most studies on spatiotemporal problems with boundaries apply simple, explicit rules like periodic boundaries \cite{mohan2020turbulence3d,guan2022turbulence2d,ross2023benchmarks}.
The studies specifically focus on boundary treatments for CNNs remain limited \cite{alguacil2021BCeffects, durand2024partial_conv}.

\citeA{zhang2023online} already tried zeroing-out land values in the input, or output, features of the whole CNN, but significant boundary artifacts are still observed, indicating that zero padding in the first layer is ineffective in mitigating the out-of-sample problem in coastal water. 
Drawing inspiration from the image processing community, which goes further and applies input filling for each layer within the network, this paper presents and compares two boundary condition treatments, i.e., zero padding and replicate padding at each layer, designed to reduce shoreline artifacts for a CNN parameterization of mesoscale eddies evaluated both offline and online.
The paper is organized as follows.
In Section \ref{sec2}, the limitations of CNN-based parameterizations near boundaries and two BC treatments in multilayer CNN models are introduced.
Section \ref{sec3} briefly introduces the CNN model used in this study, evaluating its performance with and without boundary condition treatments in an offline setting.
In Section {\ref{sec4}}, an ocean circulation model employing the ML parameterization with BC treatments is tested against an idealized wind-driven double gyre case, demonstrating the importance of boundary treatments in online CNN implementations when land points are present.
Finally, conclusions and ideas for future study are discussed in Section \ref{sec5}.

\section{Methods}\label{sec2}

In this section, we discuss the limitations of ML parameterizations using classic CNN models near ocean shorelines and introduce two straightforward yet effective methods for managing boundary conditions in multilayer CNN models.

\subsection{Limitations of CNN-based parameterizations near boundaries} \label{sec2.1}
CNNs function by sliding fixed-size kernels over images to extract spatially-local features.
Discontinuous values of a physical field across a land-sea boundary, which are not sampled in training, will lead to undetermined outputs and limit their effectiveness in environments with intricate spatial boundaries.

\begin{figure}[htbp]
\centering
\includegraphics[width=1.0\textwidth]{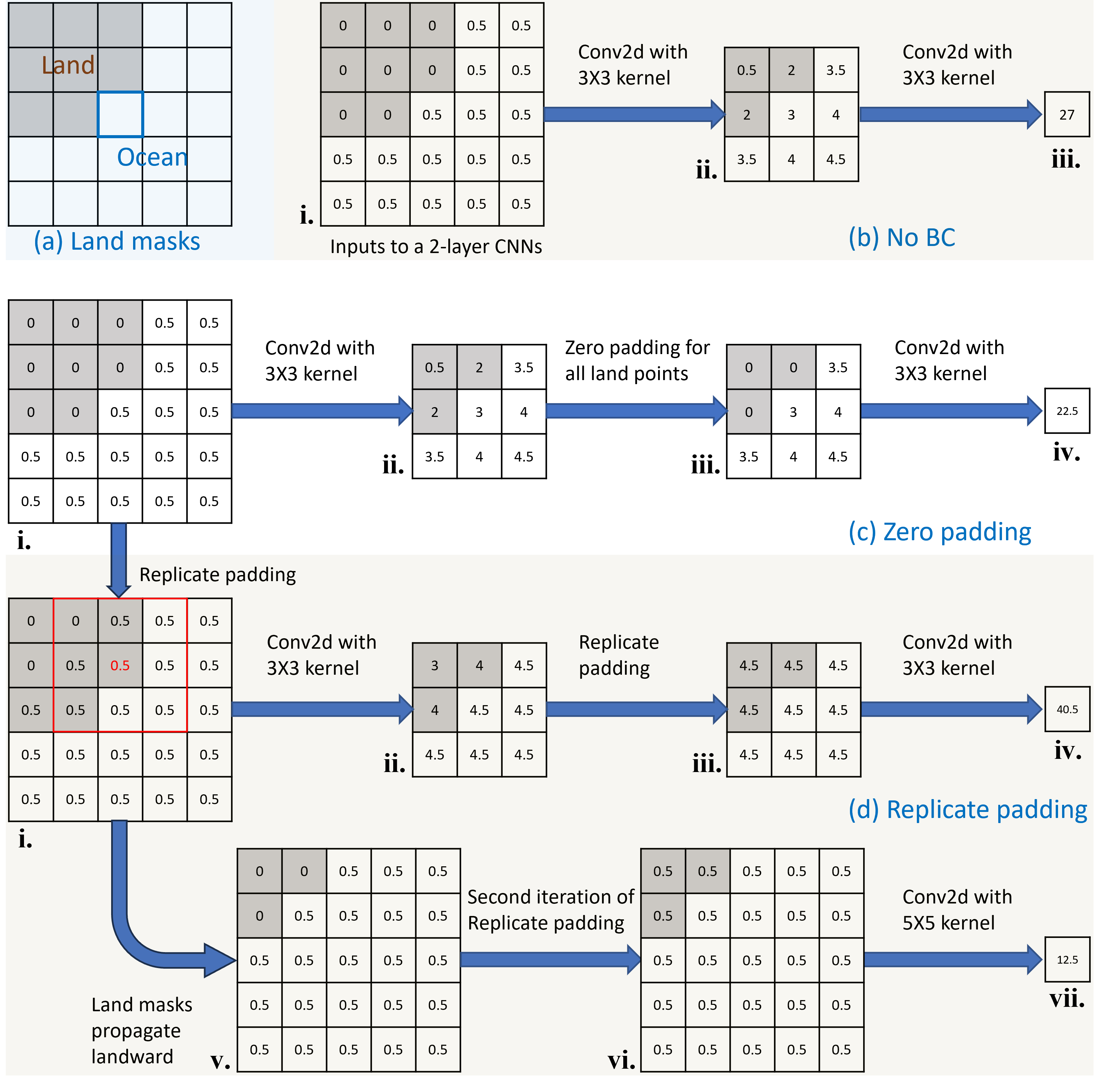}
\caption{Examples of three boundary condition strategies employed in an idealized two-layer CNN where all weights are set to 1: (a) the layout of land masks; (b) no specific treatment at land points (no padding); (c) filling land points with zeros (zero padding); (d) filling land points with values averaged from the nearest ocean points (replicate padding).}
\label{fig2_1}
\end{figure}

Consider an input data field illustrated in Figure \ref{fig2_1}(a) where the northwest portion consists of land points (grey), and the remaining points are ocean points.
In this example, the ocean points have known values of $0.5$, while the land points contain unknown (and physcially meaningless) values.
Typically, without special BC treatment, a value of zero is applied on the land points through masking of inputs (multiplying by one or zero), as shown in Figure \ref{fig2_1}(b)i.
In the case of a one-layer CNN, this masking is equivalent to setting a Dirichlet boundary condition, where an input field $y$ is set to zero at the boundary point $b$, that is $y|_b=0$.
However, merely setting the land points to zero before the first convolutional layer can introduce bias in multilayer CNN models.

Consider a two-layer CNN model, featuring two consecutive convolutional layers, each with a $3\times 3$ kernel and uniform weights of 1.
The first convolutional layer processes a $5\times 5$ input field around an ocean point (the blue cell in Figure \ref{fig2_1}(a)), and assigns values to all cells in the output $3\times3$ field including land points.
In this example, the land cells were assigned values of $0.5$, $2$, and $2$ (as indicated in the gray cells of Figure \ref{fig2_1}(b)ii).
These values in land cells do not represent any physical value or boundary conditions.
These artificial values at the land points then influence the computation of the value in the ocean point when the $3\times 3$ stencil passes the second convolutional layer (Figure \ref{fig2_1}(b)iii).
This propagating contamination from zeroed inputs highlights the need for more nuanced BC treatments in CNN-based models used for spatial parameterizations.

\subsection{Special treatments of boundary conditions in multilayer CNNs} \label{sec2.2}
To address the artifacts introduced by values at land points, we have implemented two padding strategies that incorporate information into these points at each convolutional layer in multilayer CNN models.
It should be noted that the term 'padding' used here differs from the traditional usage in CNN terminology, where 'padding' typically refers to adding values around the borders of the input images.
In our context, 'padding' specifically denotes the replacement of values at points designated as land within the land masks, which may include locations around or within the borders of the input image.

The first padding strategy we consider is "zero padding".
This is the simplest approach, where values are replaced by zeros on land points for each layer.
To illustrate, consider the scenario described in Section \ref{sec2.1}, where no boundary condition treatments were initially applied other than zero masking the land values of the input layer.
This results in the values $0.5$, $2$, and $2$ at land points after the first convolutional layer (Figures \ref{fig2_1}(b,c)ii).
These land values are reset to zero before the subsequent layer (Figure \ref{fig2_1}(c)iii), thus ensuring that a Dirichlet-like boundary condition is maintained at each convolutional layer of the CNN.

The second strategy we consider is "replicate padding", where the value at a land point is calculated as the average of values from the nearest ocean points.
This method approximates a Neumann boundary condition, where $(\partial y / \partial {\bf{n}})|_b = 0$, and ${\bf{n}}$ represents the vector normal to the boundary.
For example, the red box in Figure \ref{fig2_1}(d)i depicts a $3\times 3$ stencil used to compute the value for the central land point, averaging the values from neighboring ocean points, i.e., $0.5$ from the ocean points northeast, east, southeast, and south to the land point.
This averaging process is repeated after each convolutional layer to consistently apply a quasi-Neumann boundary condition across all layers of the CNN.

\subsection{Replicate padding for larger kernels size} \label{sec2.3}
In Section \ref{sec2.2}, we explored two padding strategies applied to sweeping a $3\times 3$ kernel over the computational domain.
In many CNN architectures, larger kernels such as $5\times 5$ and $7\times 7$ might also be employed.
For zero padding, no additional effort is required to assign a value of zero to land points encompassed by a larger kernel.

For larger kernels using replicate padding, iterations of replication are applied to fill values beyond the first layer of land points.
Figure \ref{fig2_1}(d)v-vii illustrates the mechanism of replicate padding with a $5\times 5$ kernel.
After the first application of replicate padding on the original image, the field with the first layer of replicated land points is established (Figure(d)i).
To address the values beyond these first-layer land points, we propagate the land masks landward, creating a new field of land masks (the stencil with three gray cells in Figure \ref{fig2_1}(d)v).
A second replicate padding iteration follows in Figure \ref{fig2_1}(d)vi, updating the values in the land points based on the nearest "ocean points", as indicated by the new land masks.

This iterative process of propagating land masks and updating land values is necessary when using larger kernels in CNN models.
It ensures that all land points within the kernel reach are appropriately filled, maintaining the integrity of the computational model across various kernel sizes.

\section{Offline evaluations of CNN+BC treatments}\label{sec3}
To illustrate the effectiveness of BC treatments in CNN models, we employ an existing CNN model and test it against a global dataset from a high-resolution GCM ocean simulation.
In this section, we first outline the CNN model adopted for this study.
Then we compare the offline performance of the CNN model with and without the implementation of BC treatments.

\subsection{CNN model and dataset descriptions} \label{sec3.1}
The CNN model used in this study is the stochastic-deep learning model from \citeA{Guillaumin1&Zanna-JAMES21} (hereafter referred to as GZ21).
The model was trained on the high-resolution surface horizontal velocities $\bf{u}$ from GFDL CM2.6 ocean simulations, spanning over $7,000$ days.
The surface velocities $\bf{u}$, with the nominal grid size of $1/10 ^\circ$ and sampled daily, were filtered and coarse-grained to yield ${\bar{\bfu}}$.
The subgrid momentum forcing is diagnosed as
\be
\mathbf{S} = 
(\bar{\bfu}\cdot \nabla)\bar{\bfu} - \overline{ (\bfu\cdot \nabla)\bfu }
\label{eq3.1_1}
\ee
where the overbar indicates the horizontal filtering and coarse-graining, and $\nabla$ is the horizontal gradient.
GZ21 was trained using the first $80\%$ of the data (approximately 16 years), drawn from selected four regions to represent different dynamical regimes.
The test dataset for this section comprises the remaining $20\%$ (approximately 4 years), covering the global domain.

GZ21 is structured as a fully convolutional neural network with eight convolutional layers.
The kernel sizes for the first two layers are $5 \times 5$, and $3 \times 3$ for the subsequent layers.
The layers contain 128, 64, 32, 32, 32, 32, 32, and 4 filters, respectively, each of the first 7 layers followed by ReLU activation functions.
The loss function employed is the full negative Gaussian log-likelihood, which estimates the distribution of subgrid momentum forcing given the local velocity field.
The CNN outputs both the mean and standard deviation of this distribution, $S_{C,i,j}^{(mean)}$ and $S_{C,i,j}^{(std)}$, where the stochastic subgrid momentum forcing is calculated as
\be
S_{C,i,j}=S_{C,i,j}^{(mean)}+\epsilon_{C,i,j} \cdot S_{C,i,j}^{(std)}; \hspace{0.2in}  C = x,y;\hspace{0.1in} i=1,\ldots,M;\hspace{0.1in} j=1,\ldots,N .
\label{eq3.1_2}
\ee
Here, $i$ and $j$ are the grid spatial indices, $M$ and $N$ are grid sizes in two directions, $C$ indicates the component of momentum forcing (zonal "$x$" or meridional "$y$"), and $\epsilon_{C,i,j}$ are random 2D fields sampled from the standard normal distribution, independent for each grid cell, zonal/meridional component, vertical layer, and time step.
For further details on model training and data generation, see Section 2 of \citeA{Guillaumin1&Zanna-JAMES21}.

\subsection{Metrics for offline evaluation} \label{sec3.2}
To assess the accuracy of the CNN predictions, we employ the standard root mean square error (RMSE) to measure the absolute error between the predicted values and the ground truth.
The RMSE is time-averaged at each location as follows 
\be
RMSE_{C,i,j} = \sqrt{ \frac{1}{T} \sum\limits_{t=1}^{T} \left(S_{C,i,j,t}^{(mean)}-S_{C,i,j,t}^{(true)}\right)^2 } ; \hspace{0.2in}  C = x,y
\label{eq3.2_1}
\ee
where $t$ is the time index of the snapshots and $T$ is the total number of snapshots (days) in the test dataset.
The RMSE averaged over both time and space is given by
\be
RMSE_{C} = \sqrt{ \frac{1}{M N T} \sum\limits_{i=1}^{M} \sum\limits_{j=1}^{N} \sum\limits_{t=1}^{T} \left(S_{C,i,j,t}^{(mean)}-S_{C,i,j,t}^{(true)}\right)^2 } ; \hspace{0.2in}  C = x,y
\label{eq3.2_2}
\ee

Additionally, we employ a $R^2$ coefficient, as outlined in \citeA{Guillaumin1&Zanna-JAMES21}, as a measure similar to the correlation between the predictions and the truth.
Values close to 1 signify strong predictions, while values near 0 indicate weaker predictions.
The time-averaged $R^2$ at each location is calculated as
\be
R_{C,i,j}^2 = 1- \frac{ \sum\limits_{t=1}^{T} \left(S_{C,i,j,t}^{(mean)}-S_{C,i,j,t}^{(true)}\right)^2 }{ \sum\limits_{t=1}^{T} S_{C,i,j,t}^{(true) \, 2} }; \hspace{0.2in}  C = x,y
\label{eq3.2_3}
\ee
The $R^2$ averaged over both time and space is determined by
\be
R_{C}^2 = 1- \frac{ \sum\limits_{i=1}^{M} \sum\limits_{j=1}^{N}  \sum\limits_{t=1}^{T} \left(S_{C,i,j,t}^{(mean)}-S_{C,i,j,t}^{(true)}\right)^2 }{ \sum\limits_{i=1}^{M} \sum\limits_{j=1}^{N} \sum\limits_{t=1}^{T} S_{C,i,j,t}^{(true) \, 2} }; \hspace{0.2in}  C = x,y
\label{eq3.2_4}
\ee

These metrics provide a quantitative framework for evaluating the performance of GZ21 with or without BC treatments in predicting subgrid momentum forces.

\subsection{Tests against global data} \label{sec3.3}
GZ21 was trained using data from regions devoid of land, posing a significant out-of-sample problem when predicting subgrid forces near shorelines.
GZ21 has a stencil size of $21 \times 21$ for predicting subgrid momentum forcing at an ocean grid point.
This wide stencil results in a broad coastal band (within 10 cells from the shore, approximately 60-100 km in distance when eddy-permitting resolution is applied), where the stencil includes land points, potentially affecting predictions.
In their offline evaluation on global datasets, \citeA{Guillaumin1&Zanna-JAMES21} excluded these problematic points, focusing model evaluation exclusively on open ocean areas.

\begin{figure}[htbp]
\centering
\includegraphics[width=1.0\textwidth]{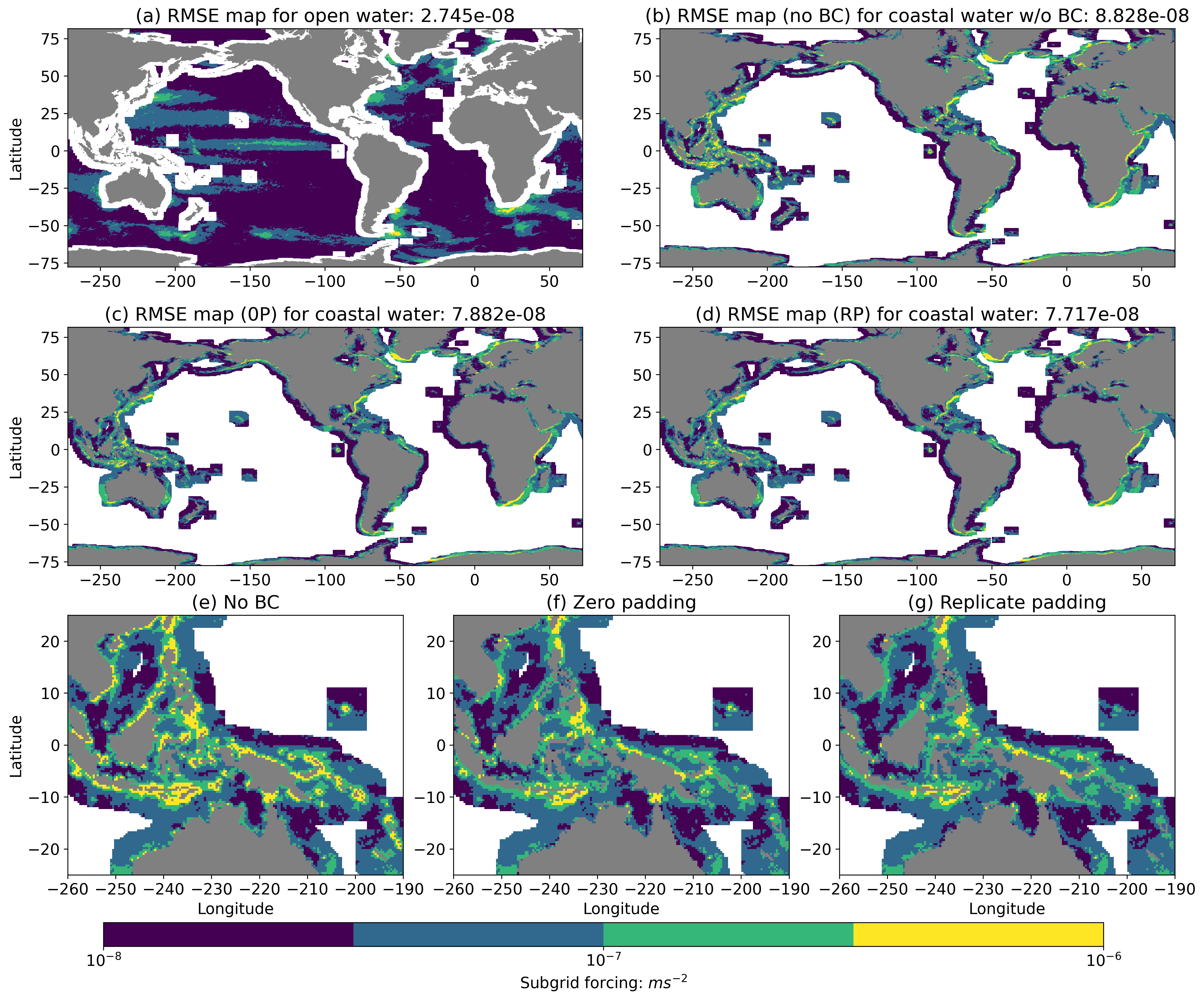}
\caption{Time-averaged global RMSE maps to true forcing for GZ21 inference of subgrid forcing $S_{x}^{(mean)}$ in open ocean domain (a), in coastal domain without special BC treatment (b), with zero padding (0P) treatment (c) and replicate padding (RP) treatment (d).
(e) to (g) are zoom-in maps of (b) to (d), respectively.
The subtitles include the RMSE averaged over both time and space.}
\label{fig3_1}
\end{figure}

In our analysis, we divide the global domain into two distinct areas: the open ocean domain and the coastal domain.
RMSE maps in Figure \ref{fig3_1} (plots a and b) illustrate the absolute prediction errors without additional BC treatments for each domain.
In this evaluation, subgrid forcing predictions in one direction are sufficiently representative of the predictions in both directions; thus, only zonal predictions are presented in this section.
The space- and time-averaged prediction errors in the coastal domain are ~3 times higher than those in the open ocean (b versus a).
The errors increase as fewer layers of coastal water points are included in the evaluation, with the errors in the layer of grid points closest to the shore being an order of magnitude higher than those in the open ocean (Table \ref{tab1}).

RMSE maps (plots c and d) in Figure \ref{fig3_1} show the result of implementing either zero padding or replicate padding which reduces the errors by approximately 11-12\%, with both strategies performing comparably.
The reduction percentage increases to about 25\% for the layer closest to the shore (Table \ref{tab1}).
These padding strategies do not alter predictions in the open ocean domain.
The changes are not clearly apparent in the global maps of RMSE but zooming into the Malay Archipelago (plots (e) to (g) of Figure \ref{fig3_1}) shows responses in ocean cells with significant numbers of land in the vicinity.
These detailed views indicate that immediate proximity to land led to larger errors without padding, and that as land cells occupy a smaller fraction of the $21 \times 21$ stencil (i.e. further away from the coast), the padding has less impact.

\begin{table}[h!]
\begin{tabular}{|c|cll|clc|clc|}
\hline
\cellcolor[HTML]{9B9B9B}{\color[HTML]{333333} }                                                                               & \multicolumn{3}{c|}{\cellcolor[HTML]{9B9B9B}{\color[HTML]{333333} No Padding}}                              & \multicolumn{3}{c|}{\cellcolor[HTML]{9B9B9B}{\color[HTML]{333333} Zero Padding}}                                                                                                            & \multicolumn{3}{c|}{\cellcolor[HTML]{9B9B9B}Replicate Padding}                                                                                                                             \\ \cline{2-10} 
\multirow{-2}{*}{\cellcolor[HTML]{9B9B9B}{\color[HTML]{333333} \begin{tabular}[c]{@{}c@{}}Number\\ of Layers\end{tabular}}} & \multicolumn{3}{c|}{\cellcolor[HTML]{9B9B9B}\begin{tabular}[c]{@{}c@{}}RMSE to\\ truth {\tiny ($10^{-7}ms^{-2}$)}\end{tabular}} & \multicolumn{2}{c|}{\cellcolor[HTML]{9B9B9B}\begin{tabular}[c]{@{}c@{}}RMSE to\\ truth {\tiny ($10^{-7}ms^{-2}$)}\end{tabular}} & \cellcolor[HTML]{9B9B9B}\begin{tabular}[c]{@{}c@{}}Improved\\ \%\end{tabular} & \multicolumn{2}{c|}{\cellcolor[HTML]{9B9B9B}\begin{tabular}[c]{@{}c@{}}RMSE to\\ truth {\tiny ($10^{-7}ms^{-2}$)}\end{tabular}} & \cellcolor[HTML]{9B9B9B}\begin{tabular}[c]{@{}c@{}}Improved\\ \%\end{tabular} \\ \hline
10                                                                                                                            & \multicolumn{3}{c|}{0.883}                                                                                  & \multicolumn{2}{c|}{7.882}                                                                                  & 10.72                                                                         & \multicolumn{2}{c|}{7.717}                                                                                 & 12.58                                                                         \\ \hline
7                                                                                                                             & \multicolumn{3}{c|}{1.029}                                                                                  & \multicolumn{2}{c|}{0.905}                                                                                  & 12.05                                                                         & \multicolumn{2}{c|}{0.883}                                                                                 & 14.19                                                                         \\ \hline
5                                                                                                                             & \multicolumn{3}{c|}{1.188}                                                                                  & \multicolumn{2}{c|}{1.028}                                                                                  & 13.47                                                                         & \multicolumn{2}{c|}{0.999}                                                                                 & 15.88                                                                         \\ \hline
4                                                                                                                             & \multicolumn{3}{c|}{1.305}                                                                                  & \multicolumn{2}{c|}{1.114}                                                                                  & 14.64                                                                         & \multicolumn{2}{c|}{1.081}                                                                                 & 17.16                                                                         \\ \hline
3                                                                                                                             & \multicolumn{3}{c|}{1.472}                                                                                  & \multicolumn{2}{c|}{1.232}                                                                                  & 16.30                                                                         & \multicolumn{2}{c|}{1.192}                                                                                 & 19.02                                                                         \\ \hline
2                                                                                                                             & \multicolumn{3}{c|}{1.734}                                                                                  & \multicolumn{2}{c|}{1.404}                                                                                  & 19.03                                                                         & \multicolumn{2}{c|}{1.358}                                                                                 & 21.69                                                                         \\ \hline
1                                                                                                                             & \multicolumn{3}{c|}{2.209}                                                                                  & \multicolumn{2}{c|}{1.677}                                                                                  & 24.08                                                                         & \multicolumn{2}{c|}{1.634}                                                                                 & 26.03                                                                         \\ \hline
\end{tabular}
\caption{Offline evaluation results by inferring the subgrid forcing $S_{x}^{(mean)}$ using GZ21 in the coastal domain with different BC padding strategies, based on the metric of RMSE averaged over both time and space, as well as the improved percentage of using these BC padding strategies. 'Number of Layers' refers to the number of layers of water points near coastlines that are included in the evaluation.}
\label{tab1}
\end{table}

\begin{figure}[htbp]
\centering
\includegraphics[width=1\textwidth]{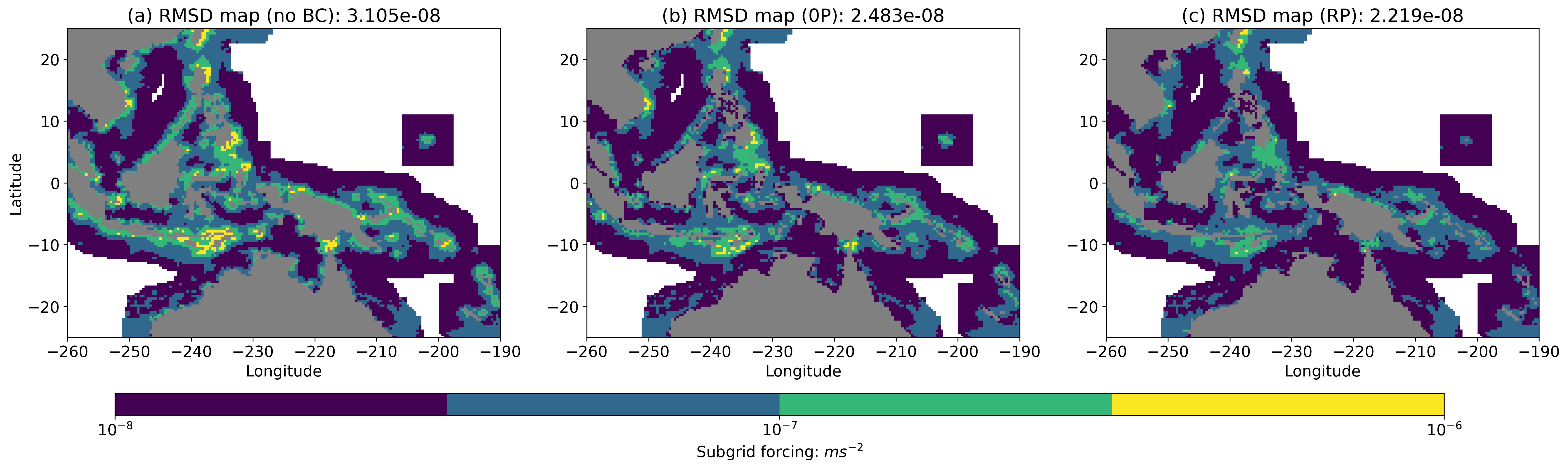}
\caption{Time-averaged $RMSD^{(\mathrm{model})}$ maps, focusing on the Malay Archipelago, of subgrid forcing $S_{x}^{(mean)}$ between inference of the original CNN model GZ21 and the retrained CNN model GZ21-T2 with global data (a) without special BC treatment, (b) with zero padding treatment, and (c) replicate padding treatment. The RMSD values in subtitles are global averaging over both time and space.}
\label{fig3_2}
\end{figure}

The significant errors observed in coastal domain predictions are a manifestation of the out-of-sample problem.
The particular predictions in the coastal domain should vary from training to training due to the random initialization of weights.
To check this, we retrained a new CNN model, GZ21-T2, following the exact procedures described in \citeA{Guillaumin1&Zanna-JAMES21}.
Figure \ref{fig3_2} compares the root mean square differences (RMSD) for coastal domain predictions between GZ21 and GZ21-T2, calculated as
\be
RMSD^{(\mathrm{model})}_{x,i,j} =  \sqrt{ \frac{1}{T} \sum\limits_{t=1}^{T} \left(S_{x,i,j,t}^{(\mathrm{GZ21}, mean)}-S_{x,i,j,t}^{(\mathrm{GZ21-T2}, mean)}\right)^2 } ; \hspace{0.2in}
\label{eq3.3_1}
\ee
with and without BC treatments.
The results indicate that BC treatments effectively reduce the randomness of out-of-sample predictions, indicated by the notably smaller overall differences when BC treatments are applied.

\begin{figure}[htbp]
\centering
\includegraphics[width=0.8\textwidth]{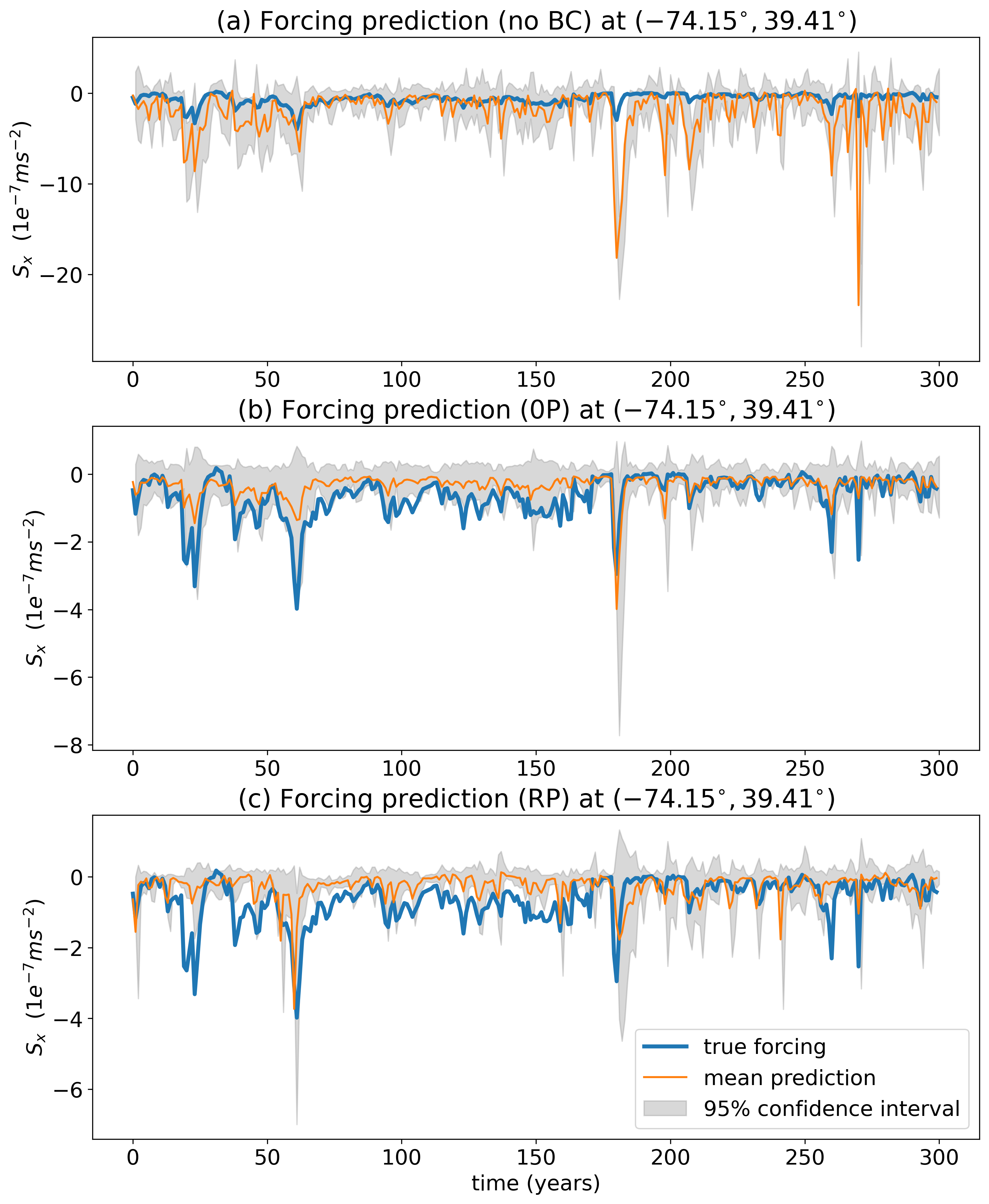}
\caption{Time series of true forcing (blue), of the mean predictions ($S_{x}^{(mean)}$, orange), and of the 95\% confidence interval
($\pm 1.96 \, S_{x,i,j}^{(std)}$, shaded in grey) at a location $(-74.15^{\circ},39.41^{\circ})$ near Atlantic City, New Jersey.}
\label{fig3_4}
\end{figure}

The $R^2$ coefficient, as described in Section \ref{sec3.2}, serves as an effective indicator of the overall performance of ML models within the area of interest.
The global $R^2$ values averaged over both time and space (Eq. \ref{eq3.2_4}) are 0.381 for GZ21 without BC treatments, 0.563 for GZ21 with zero paddings, and 0.558 for GZ21 with replicate padding.
The variability in $R^2$ averaged in space reflects varying levels of bias from out-of-sample predictions near the coasts under different BC treatments.
For example, at a location $(-74.15^{\circ},39.41^{\circ})$ near Atlantic City, New Jersey, GZ21 tends to predict significantly higher absolute values.
Figure \ref{fig3_4}(a) contrasts the predicted forcing from GZ21 with the true forcing, where the blue lines represent true forcing, the orange lines represent the GZ21 predictions, and the dashed green lines represent the 95\% confidence interval.
The true forcing is in the range of $[-4,0]\times10^{-7} ms^{-2}$, while the mean part of the forcing from the GZ21 predictions is in the range of $[-20,0]\times10^{-7} ms^{-2}$.
The BC treatments can effectively reduce the prediction range, and both strategies reduce the range of forcing to $[-4,0]\times10^{-7} ms^{-2}$.
These plots highlight the efficacy of BC treatments in narrowing the prediction range to more closely align with the actual measurements, despite not perfectly matching the truth.


To verify the reproducibility of GZ21, we repeated the training process 6 additional times (GZ21-T3 to GZ21-T8).
It is important to note that each model may exhibit significant performance variations in coastal domains (out-of-sample predictions) due to the random initialization of weights during each training session.
Table \ref{tab2} lists the global $R^2$ coefficients, coastal RMSE between predictions and truth, and coastal RMSD between different model predictions, both with and without BC treatments.
The standard deviation of the $R^2$ column for no padding, zero padding, and replicate padding are 0.0179, 0.0049, 0.0056, respectively.
The significantly lower numbers from the offline evaluations with BC treatments confirm the improved reproducibility of GZ21 with BC treatments compared to GZ21 without BC treatments.

\begin{table}[h!]
\resizebox{\columnwidth}{!}{%
\begin{tabular}{|c|ccc|ccc|ccc|}
\hline
\rowcolor[HTML]{9B9B9B} 
\cellcolor[HTML]{9B9B9B}{\color[HTML]{333333} }                                                                        & \multicolumn{3}{c|}{\cellcolor[HTML]{9B9B9B}{\color[HTML]{333333} No Padding}}                                                                                                                                                  & \multicolumn{3}{c|}{\cellcolor[HTML]{9B9B9B}{\color[HTML]{333333} Zero Padding}}                                                                                                                                                              & \multicolumn{3}{c|}{\cellcolor[HTML]{9B9B9B}Replicate Padding}                                                                                                                                                                                \\ \cline{2-10} 
\rowcolor[HTML]{9B9B9B} 
\multirow{-2}{*}{\cellcolor[HTML]{9B9B9B}{\color[HTML]{333333} \begin{tabular}[c]{@{}c@{}}Model \\ Name\end{tabular}}} & \multicolumn{1}{c|}{\cellcolor[HTML]{9B9B9B}$R^2$} & \multicolumn{1}{c|}{\cellcolor[HTML]{9B9B9B}\begin{tabular}[c]{@{}c@{}}{\small RMSE to} \\ {\small truth} {\tiny ($10^{-8}ms^{-2}$)}\end{tabular}} & \begin{tabular}[c]{@{}c@{}}{\small RMSD to} \\ {\small GZ21} {\tiny ($10^{-8}ms^{-2}$)}\end{tabular} & \multicolumn{1}{c|}{\cellcolor[HTML]{9B9B9B}$R^2$} & \multicolumn{1}{c|}{\cellcolor[HTML]{9B9B9B}\begin{tabular}[c]{@{}c@{}}{\small RMSE to} \\ {\small truth} {\tiny ($10^{-8}ms^{-2}$)}\end{tabular}} & \cellcolor[HTML]{9B9B9B}\begin{tabular}[c]{@{}c@{}}{\small RMSD to} \\ {\small GZ21} {\tiny ($10^{-8}ms^{-2}$)}\end{tabular} & \multicolumn{1}{c|}{\cellcolor[HTML]{9B9B9B}$R^2$} & \multicolumn{1}{c|}{\cellcolor[HTML]{9B9B9B}\begin{tabular}[c]{@{}c@{}}{\small RMSE to} \\ {\small truth} {\tiny ($10^{-8}ms^{-2}$)}\end{tabular}} & \cellcolor[HTML]{9B9B9B}\begin{tabular}[c]{@{}c@{}}{\small RMSD to} \\ {\small GZ21} {\tiny ($10^{-8}ms^{-2}$)}\end{tabular} \\ \hline
GZ21                                                                                                                   & \multicolumn{1}{c|}{0.381}                      & \multicolumn{1}{c|}{8.828}                                                                                   & /                                                              & \multicolumn{1}{c|}{0.563}                      & \multicolumn{1}{c|}{7.882}                                                                         & /                                                                                      & \multicolumn{1}{c|}{0.558}                      & \multicolumn{1}{c|}{7.717}                                                                         & /                                                                                      \\ \hline
GZ21-T2                                                                                                               & \multicolumn{1}{c|}{0.385}                      & \multicolumn{1}{c|}{8.861}                                                                                   & 3.105                                                          & \multicolumn{1}{c|}{0.559}                      & \multicolumn{1}{c|}{7.783}                                                                         & 2.483                                                                                  & \multicolumn{1}{c|}{0.546}                      & \multicolumn{1}{c|}{7.716}                                                                         & 2.219                                                                                  \\ \hline
GZ21-T3                                                                                                               & \multicolumn{1}{c|}{0.401}                      & \multicolumn{1}{c|}{8.846}                                                                                   & 3.210                                                          & \multicolumn{1}{c|}{0.557}                      & \multicolumn{1}{c|}{7.823}                                                                         & 2.617                                                                                  & \multicolumn{1}{c|}{0.540}                      & \multicolumn{1}{c|}{7.772}                                                                         & 2.302                                                                                  \\ \hline
GZ21-T4                                                                                                               & \multicolumn{1}{c|}{0.419}                      & \multicolumn{1}{c|}{8.635}                                                                                   & 2.969                                                          & \multicolumn{1}{c|}{0.572}                      & \multicolumn{1}{c|}{7.781}                                                                         & 2.505                                                                                  & \multicolumn{1}{c|}{0.545}                      & \multicolumn{1}{c|}{7.736}                                                                         & 2.244                                                                                  \\ \hline
GZ21-T5                                                                                                               & \multicolumn{1}{c|}{0.367}                      & \multicolumn{1}{c|}{8.874}                                                                                   & 2.990                                                          & \multicolumn{1}{c|}{0.561}                      & \multicolumn{1}{c|}{7.832}                                                                         & 2.472                                                                                  & \multicolumn{1}{c|}{0.547}                      & \multicolumn{1}{c|}{7.748}                                                                         & 2.193                                                                                  \\ \hline
GZ21-T6                                                                                                               & \multicolumn{1}{c|}{0.400}                      & \multicolumn{1}{c|}{8.679}                                                                                   & 3.062                                                          & \multicolumn{1}{c|}{0.558}                      & \multicolumn{1}{c|}{7.760}                                                                         & 2.458                                                                                  & \multicolumn{1}{c|}{0.546}                      & \multicolumn{1}{c|}{7.707}                                                                         & 2.183                                                                                  \\ \hline
GZ21-T7                                                                                                               & \multicolumn{1}{c|}{0.367}                      & \multicolumn{1}{c|}{8.938}                                                                                   & 3.110                                                          & \multicolumn{1}{c|}{0.558}                      & \multicolumn{1}{c|}{7.891}                                                                         & 2.558                                                                                  & \multicolumn{1}{c|}{0.540}                      & \multicolumn{1}{c|}{7.790}                                                                         & 2.220                                                                                  \\ \hline
GZ21-T8                                                                                                               & \multicolumn{1}{c|}{0.395}                      & \multicolumn{1}{c|}{8.793}                                                                                   & 3.042                                                          & \multicolumn{1}{c|}{0.559}                      & \multicolumn{1}{c|}{7.878}                                                                         & 2.532                                                                                  & \multicolumn{1}{c|}{0.546}                      & \multicolumn{1}{c|}{7.767}                                                                         & 2.262                                                                                  \\ \hline
\end{tabular}
}
\caption{Offline evaluation results by inferring the subgrid forcing $S_{x}^{(mean)}$ using original CNN model GZ21 or retrained model GZ21-T2 to GZ21-T8 with different BC padding strategies, based on two metrics of global $R^2$ and coastal RMSE averaged over both time and space, as well as coastal RMSD of the forcing prediction between retrained models and GZ21.}
\label{tab2}
\end{table}

\section{Online implementations of CNN+BC with MOM6}\label{sec4}
The ultimate goal of developing ML parameterizations is to improve the online solution of numerical models.
While the overall offline performance is excellent in Section \ref{sec3}, it does not assure comparable online success \cite{Ross-et-al2022}.
When these parameterizations are incorporated into a coarse-resolution ocean model and executed over extended periods, local errors introduced by the parameterization can accumulate and/or spread throughout the simulation.
This can subsequently contaminate the entire domain or, in severe cases, cause the simulation to fail.

In this section, we further explore the effectiveness of boundary condition (BC) treatments within the GZ21 parameterization for online inference.
We test the model both with and without BC treatments in an idealized case for which we can afford to run a fine-resolution "truth": a wind-driven double gyre \cite{hallberg_rhines_2000}.
We will first briefly outline the ocean model used for this study and the setup of the case.
Then we will examine the evaluation results and discuss the computational costs associated with implementing BC treatments in online simulations.

\subsection{Ocean model and case setup} \label{sec4.1}
The numerical model employed in this study is the Modular Ocean Model version 6 (MOM6) \cite{adcroft2019gfdl}, the ocean component of the NOAA coupled global climate and earth system models developed at GFDL.
We apply MOM6 under the assumption of an adiabatic limit with no buoyancy forcing, which simplifies the ocean dynamics into the stacked shallow water equations.
This assumption facilitates the testing of the ML parameterization of subgrid momentum forcing (Eq. \ref{eq3.1_1}) in an idealized setting of a primitive equation model.
The governing equations are discretized on a C-type staggered grid, positioning the velocity components ($\bar{\bfu}$ in Eq. \ref{eq3.1_1}) on cell faces and the subgrid forcing ($\mathbf{S}$ in Eq. \ref{eq3.1_1}) at the cell centers.
During the implementation of the ML parameterization, velocity components from MOM6 are interpolated to the cell centers and then used as inputs to the ML model to infer subgrid forcing, which is then interpolated back to the cell faces.
Further details on the model descriptions are available in Section 2.1 of \citeA{zhang2023online}.

For this study, the ocean model is configured to simulate two idealized wind-driven double gyre scenarios.
The first configuration (hereafter referred to as C1) features a bowl-shaped basin \cite{hallberg_rhines_2000} extending from $0^\circ$ to $22^\circ$ in longitude and from $30^\circ$ to $50^\circ$ in latitude, with a depth ranging from $-2000$m to $0$m in vertical.
A vertical wall is placed at the southern boundary.
The flow includes two vertical layers with constant water density in each layer, and no computations involving equation of state, temperature and salinity.
The circulation is driven by wind and balanced by bottom friction.
The simulations start from rest and continue for a duration of 10 years.
Further details can be found in Section 3.1 of \citeA{zhang2023online}.

The second configuration (hereafter referred to as C2) introduces a box-shaped island in the center of the domain based on the first configuration, located between $8.5^\circ$ to $13.5^\circ$ in longitude and $37.5^\circ$ to $42.5^\circ$ in latitude \cite<see Figure 17 in >{zhang2023online}.
This island represents a significant topographic obstacle in the path of the wind-driven jet and we expect the abrupt nature of the obstacle to test the limits of the ML parameterization near boundaries.

The evaluations are conducted using a coarse grid model with $1/4^\circ$ horizontal resolution (hereafter referred to as R4), which is "eddy-permitting" but not fine enough to resolve all mesoscale eddy dynamics.
By applying the ML parameterization in R4 (hereafter referred to as R4-P), we compare its performance to that of a fine grid model with a $1/32^\circ$ horizontal resolution (hereafter referred to as R32), which is capable of fully capturing mesoscale eddy processes.
In this section, the total Kinetic Energy (KE) of the flow is the only metric used to quantitatively evaluate the online performance of the CNN model.


\subsection{Results with various BC treatments} \label{sec4.2}
\begin{figure}[htbp]
\centering
\includegraphics[width=1.0\textwidth]{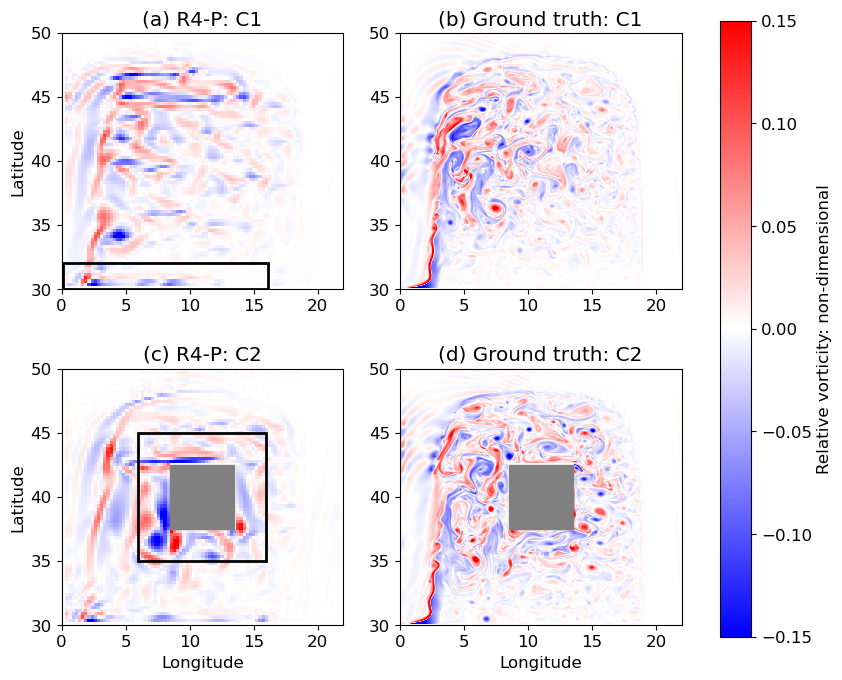}
\caption{Snapshots of the upper layer relative vorticity at the end of C1 and C2 simulations from the coarse resolution model with ML parameterizations R4-P (a,c) and the fine resolution R32 (b,d).
The ML parameterizations do not use special BC treatments.
The black rectangle indicates the region where the unrealistic eddies are generated.
}
\label{fig4_1}
\end{figure}

The study of \citeA{zhang2023online} found that when GZ21 is applied in MOM6 to predict the subgrid mesoscale momentum forcing near boundaries, it generates artifacts that cause the over-energization of the flow.
The structures highlighted within black rectangles in Figure \ref{fig4_1} illustrate the typical artifacts, which are absent in the higher-resolution model R32.

\begin{figure}[htbp]
\centering
\includegraphics[width=1.0\textwidth]{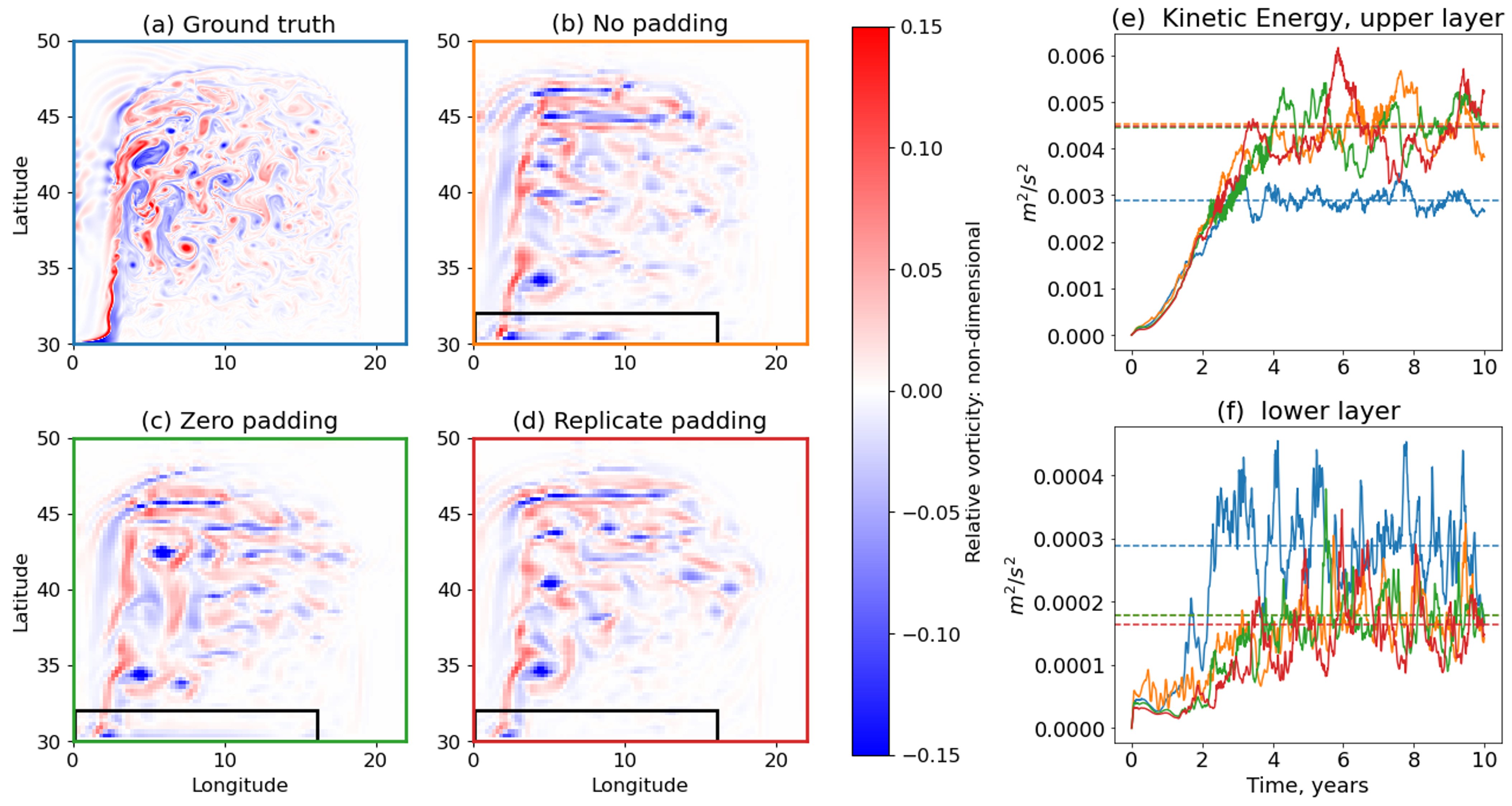}
\caption{(a)-(d): Snapshots of the upper layer relative vorticity at the end of C1 simulations from: (a) the fine resolution R32; (b) the coarse resolution model using ML parameterization R4-P without special BC treatment; (c) with zero padding strategy; or (d) with replicate padding.
(e) \& (f): Comparison of KE time series for the flow upper layer and lower layer between the four simulations from (a) to (d).
The dashed lines are the time-mean values of KE over the last 5 years and the colors of solid lines correspond to the edge colors of plots (a) to (d), where blue is for ground truth, orange for no padding, green for zero padding and red for replicate padding. The black rectangle indicates the region where the unrealistic eddies are generated.}
\label{fig4_2}
\end{figure}

We first consider online evaluations of the original GZ21 model (not retrained) but with different BC treatment during inference.
Figure \ref{fig4_2}(a-d) shows the relative vorticity snapshots of the upper flow from R32, R4-P without padding, R4-P with zero padding, and R4-P with zero and replicate padding strategies in the C1 scenario.
Figure \ref{fig4_2}(e,d) compares the time series of KE in both the upper and lower layer flows under different padding strategies, with line colors corresponding to the edges of each plot from (a) to (d).
\citeA{zhang2023online} demonstrated that the parameterizations based on GZ21 tend to over-energize the flow in the upper layer while under-energizing the flow in the lower layer.
Neither of the two padding strategies affect the overall energy injection in each layer, but they are effective in mitigating artifacts near the southern boundary.
In C1, the performance of both treatments appears similar.

\begin{figure}[htbp]
\centering
\includegraphics[width=1.0\textwidth]{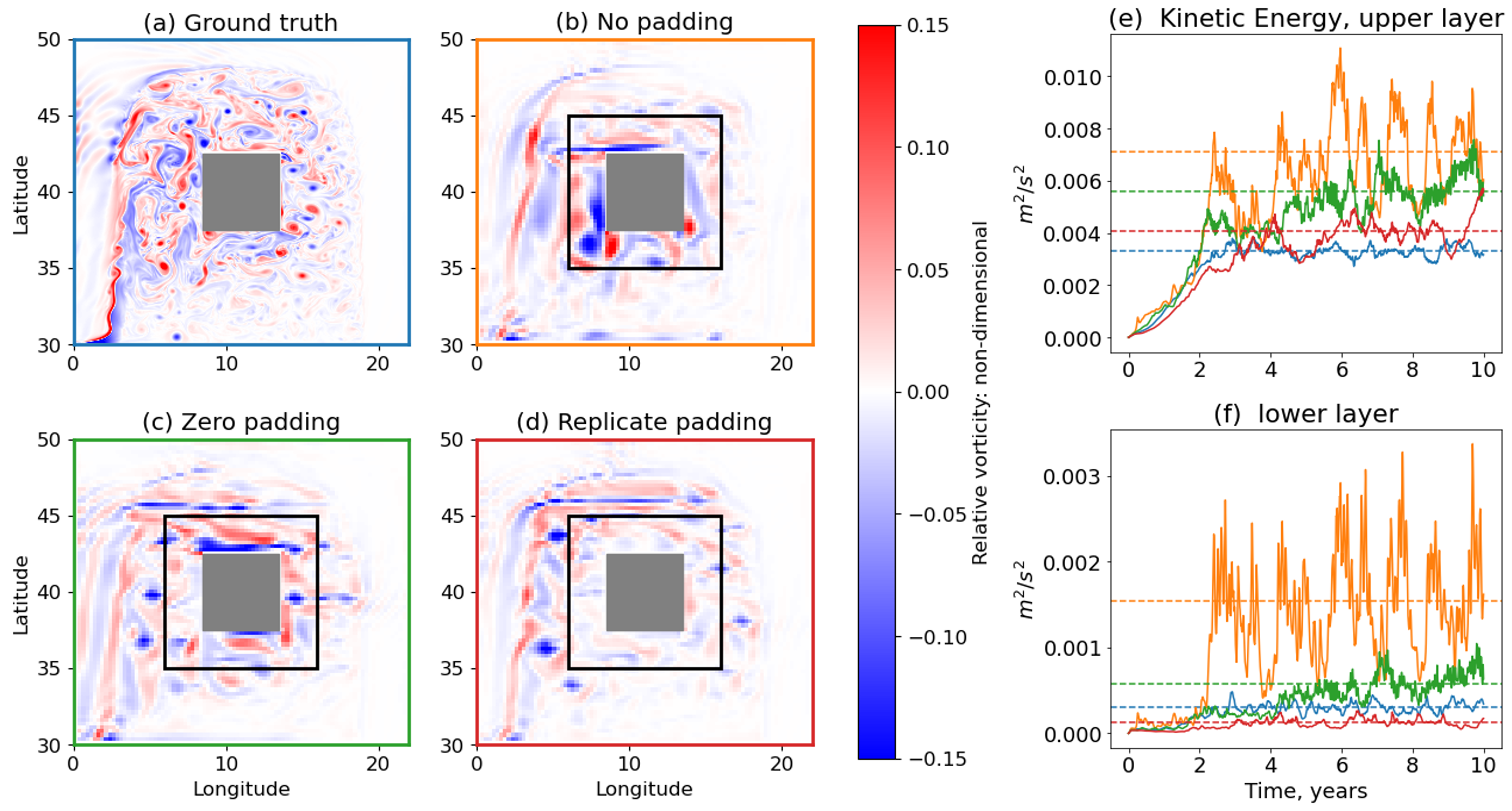}
\caption{Same to Figure \ref{fig4_2} but tested on C2.}
\label{fig4_3}
\end{figure}

The ML parameterizations were also tested in C2.
Figure \ref{fig4_3}(a-d) shows the relative vorticity snapshots from the ground truth and ML parameterizations with different BC treatments.
Without special BC treatments, obvious sheared structures appear both around the box island, as well as at the southern boundary as we observe in C1 (Figure \ref{fig4_3}(b)).
Zero padding removes the relatively weak artifacts near the southern boundary but does not affect the strong sheared structures around the island (Figure \ref{fig4_3}(c)).
In contrast, replicate padding effectively eliminates the artifacts in both boundary regions, aligning the better relative vorticity snapshot with the ground truth.
This is also evident in the time series of KE (Figure \ref{fig4_3}(e,f)), where no treatment results in a more energetic flow, zero padding reduces energy from artifacts, and replicate padding closely matches the KE to that in the ground truth.
Thus, replicate padding is the most effective approach considered in C2.
It should be noted that the zonal elongation of eddies is observed in Figures \ref{fig4_2}(b-d) and \ref{fig4_3}(b-d). We hypothesize that the four regions selected in GZ21 exhibit a tendency towards zonal flows, as discussed in Section 4.4 of \citeA{zhang2023online}. The discussion of this issue is beyond the scope of this paper.

\begin{figure}[htbp]
\centering
\includegraphics[width=1.0\textwidth]{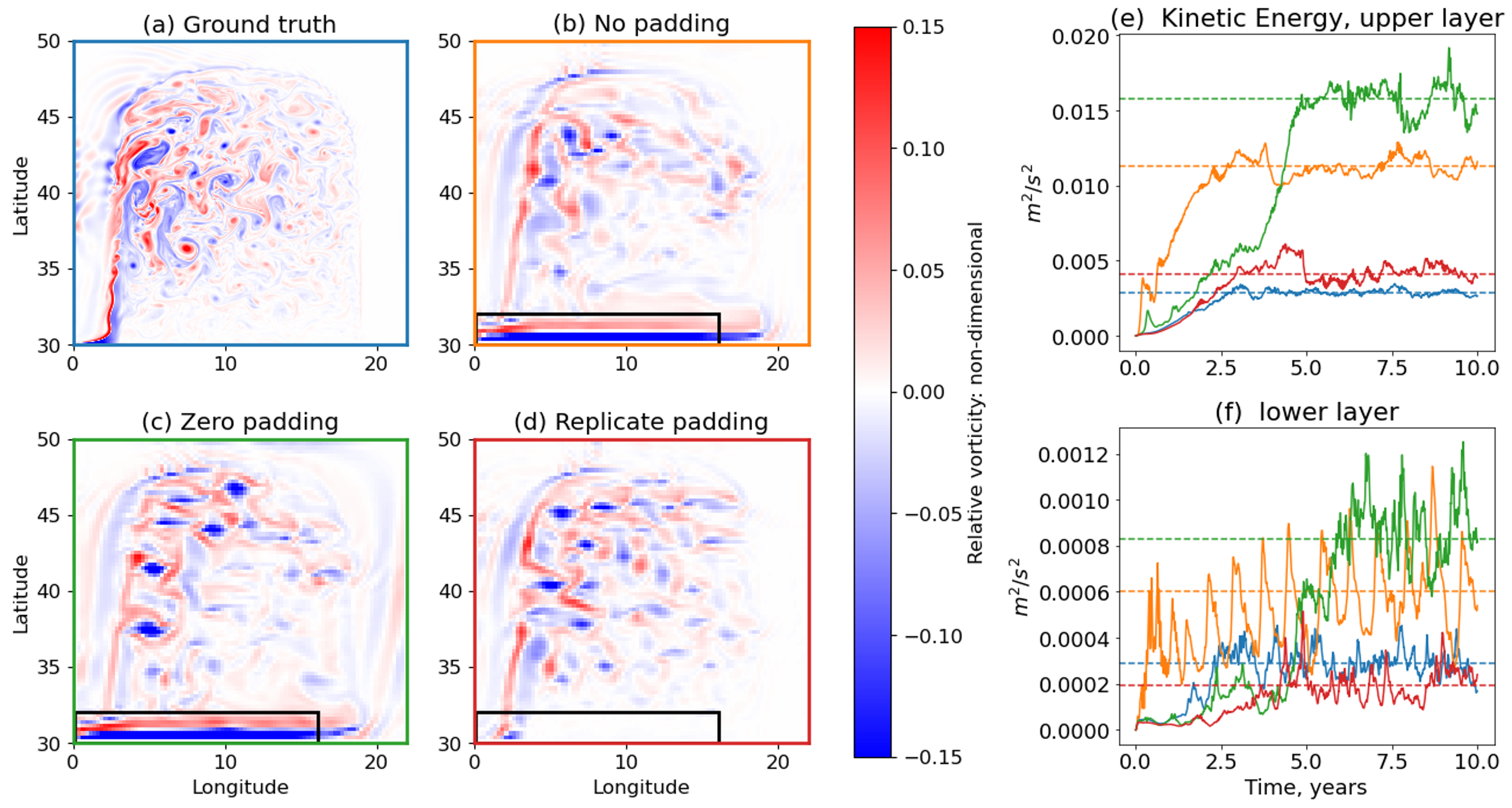}
\caption{Same to Figure \ref{fig4_2} but tested using GZ21-T2 for the ML parameterizations.}
\label{fig4_4}
\end{figure}

\begin{figure}[htbp]
\centering
\includegraphics[width=1.0\textwidth]{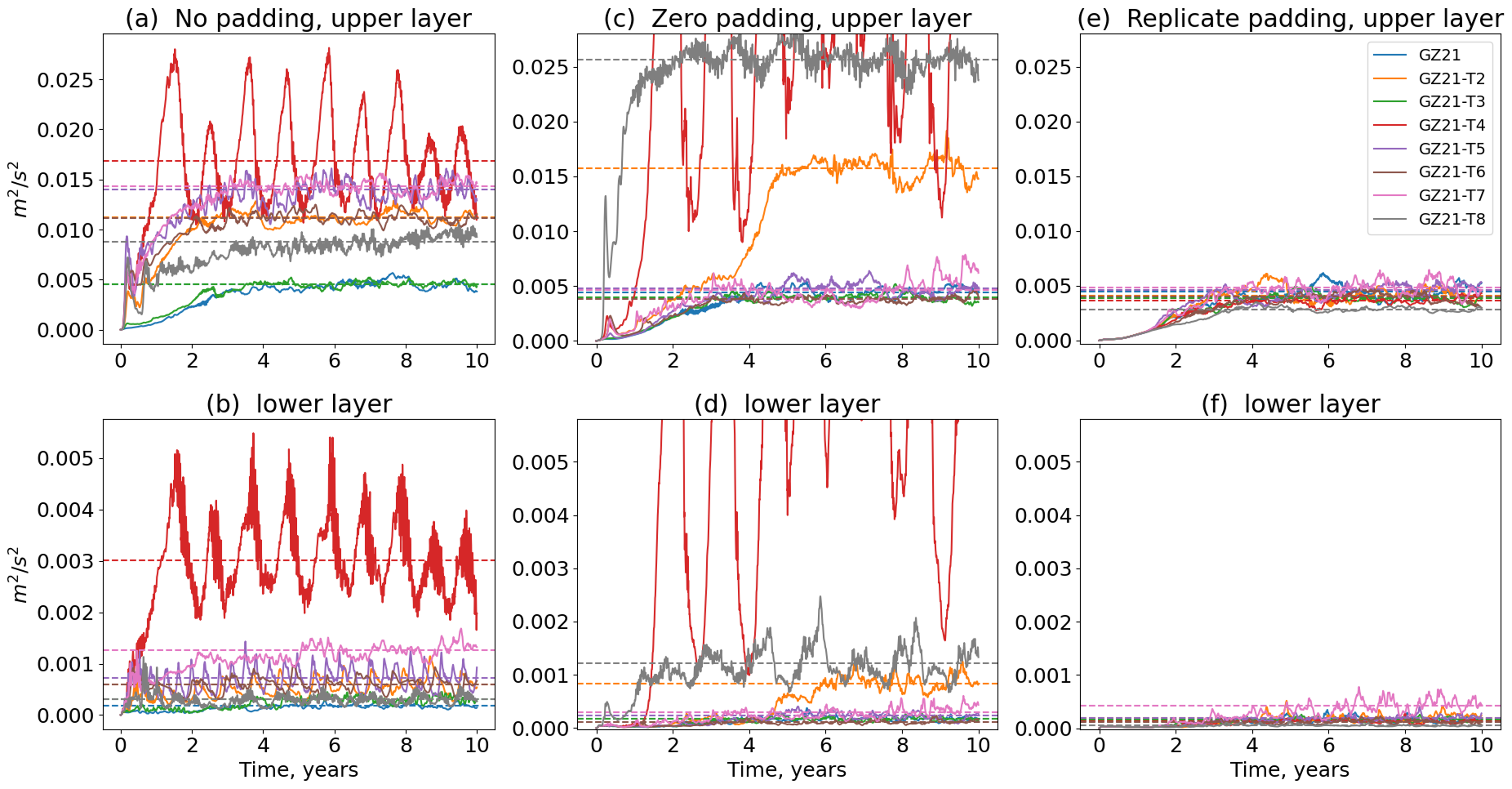}
\caption{Comparison of KE time series for the flow upper layer and lower layer between the simulations using GZ21 and retrained CNN models.
The first column of plots (a) and (b) represent the simulation results without any BC treatment, the second column of plots (c) and (d) represent the results with zero padding BC treatment, and the third column of plots (e) and (f) represent the results with replicate padding BC treatment.
The dashed lines are the time-mean values of KE over the last 5 years and each row shares the same axis range to better compare the results with different BC strategies.}
\label{fig4_8}
\end{figure}

We now consider whether the padding strategy can reduce the sensitivity of network performance to the random initialization of weights in training processes; the GZ21-T2 model is retrained with the same data as GZ21 and without padding used during training (just like what GZ21 does), but reevaluated with the various padding treatments.
Figure \ref{fig4_4} shows results with the retrained model in C1.
It is interesting to see that the zero padding strategy fails to eliminate the artifacts near the southern boundary (Figure \ref{fig4_4}(c)) as it does with GZ21.
After a 5-year run, the flow becomes even more energetic compared to simulations that did not use any BC treatments.
In contrast, replicate padding with GZ21-T2 performs well in C1 for artifact eliminations.
Tests are also conducted on configuration C2; however, simulations using GZ21-T2 with no padding and zero padding failed in C2, as excessive energy was injected into the flow, causing the models to blow up.
This issue of over-energization near boundaries is not isolated to GZ21-T2; some other retrained models (models in Table \ref{tab2}) also struggle to eliminate artifacts using the zero padding method.
The KE time series depicted in Figure \ref{fig4_8}(c,d) demonstrate that using zero padding in GZ21-T4 and GZ21-T8 similarly leads to poorer predictions of energy injection in configuration C1.
In contrast, the replicate padding BC treatment consistently reduces energy injection to a more reasonable range across all retrained models, aligning more closely with the ground truth.

In addition to the effectiveness of the BC strategies, their computational cost is also a critical consideration.
The zero padding strategy does not significantly increase the computational load compared to the approach without BC treatments.
Implementing zero padding across each CNN convolutional layer for GZ21 results in approximately a 10\% increase in wall clock time for CNN inference.
In contrast, the process of filling the nearest value to the land points is considerably more costly due the stencil operations needed to propagate values. 
We tried optimizing this operation as a pre-computed sparse matrix, which significantly improved performance, yet only to within a factor of 2 compared to the no-padding model.
While this cost represents a substantial increase compared to the computational costs associated with no padding and zero padding, it renders the strategy feasible.
Furthermore, optimizations in sparse matrix operations could potentially further reduce the time required for inference.

\section{Conclusions}\label{sec5}

This study was motivated by a notable issue identified in the previous study of \citeA{zhang2023online}, where distinct artifacts near boundaries were observed in the online implementation of the ML parameterization based on GZ21.
These artifacts were suspected to be generated by out-of-sample predictions near boundaries because GZ21 was only trained in the open ocean.
Developing a new network capable of accommodating the complex flow regimes near intricate shorelines likely requires substantial effort and sophisticated architectures with uncertain convergence.
We instead explored the use of specialized BC treatments to use with existing CNN models to address over-energization caused by out-of-sample predictions near coasts.

Our offline evaluations of the existing CNN model GZ21 with the global dataset CM2.6, demonstrated that both zero padding and replicate padding strategies can effectively reduce RMSE near coastlines.
The significant RMSE in coastal domains is primarily due to the unconstrained out-of-sample predictions in these regions with the original GZ21.
For example, force predictions near Atlantic City, New Jersey, indicated that GZ21 without padding tends to produce a wide range of values, whereas GZ21 with zero or replicate padding can mitigate this randomness and narrow the range of out-of-sample predictions.
In Addition, it is important to note that the unconstrained out-of-sample error varies significantly among GZ21 and the retained models of GZ21 due to the random initialization of weights during their training processes. 
Implementing boundary treatments can help effectively restrict the error for all models within a reasonable range.

Online evaluations of two configurations, the wind-driven double gyre (C1) and double gyre with in island in the center (C2), confirmed that using the replicate padding strategy as a BC treatment can effectively eliminate boundary artifacts in the online implementation of ML parameterizations, outperforming both the no padding and zero padding approaches.
Our reproducibility tests indicated that GZ21 was fortuitously trained such that its predictions near boundaries do not generate overly strong sheared artifacts, which would lead to excessive energy within the flow or even simulation blowup.
Although the no padding strategy was effective in offline evaluations and for several retrained models in this online evaluation, it was not universally useful, whereas the replicate padding strategy proved effective at avoiding artifacts across all retrained models in this study.

The objective of this study is to develop strategies that limit prediction errors in regions where the ML parameterizations lack sufficient 'knowledge'.
Typically, ML parameterizations for ocean subgrid forcing neglect consideration of land points during training due to the complexities of coastal regions and the intricacies of managing land point values.
This is true for many conventional parameterizations also.
In essence, employing replicate padding in a CNN model for coastal regions minimizes implied gradients near coasts, which naturally reduces the magnitude of predictions from the model.
This approach offers a viable pathway whereby an existing CNN model trained on open water data can be used to predict forces in coastal areas without generating strongly anomalous outputs.
This capability is crucial in online implementations of ML parameterizations because any extreme value introduced by CNN inference can eventually propagate throughout the domain, contaminating the solution, or even leading to failure of the simulation.


%
%
%
%

%
%

%

%

\section*{Open Research}
The source code of the MOM6 version used for implementing the ML parameterization is accessible through Zenodo \cite{MOM6Forpy_src}, while the CNN model files used for the online evaluation in this study (GZ21) can also be accessed via Zenodo \cite{ForpyCNNGZ21}. The files for offline global evaluations can be accessed via Zenodo \cite{GZ21_model}. To facilitate the setup process for the wind-driven double gyre case in the study, we have made the setup files available online \cite{doublegyre_setup}. 

\acknowledgments
We thank all members of the M$^2$LInES team for helpful discussions and their support throughout this project. We thank Qian Shao and Wenda Zhang for useful comments on a draft of this manuscript, and Arthur Guillaumin for assistance with the networks.
This research received support through Schmidt Sciences, LLC.
AA was also supported by award NA18OAR4320123, from the National Oceanic and Atmospheric Administration (NOAA), U.S. Department of Commerce and which funded the Princeton Stellar computer resources used for the inference stage of the research.
The statements, findings, conclusions, and recommendations are those of the author(s) and do not necessarily reflect the views of the National Oceanic and Atmospheric Administration, or the U.S. Department of Commerce.
This research was also supported in part through the NYU IT High Performance Computing resources, services, and staff expertise.


%
%



\bibliography{zhang-etal-james24}

%
%
%
%
%

\end{document}